# Interface boundary conditions for dynamic magnetization and spin wave dynamics in a ferromagnetic layer with the interface Dzyaloshinskii-Moriya interaction

M. Kostylev

*School of Physics, M013, University of Western Australia, Crawley 6009, Western Australia, Australia*

Abstract: In this work we derive the interface exchange boundary conditions for the classical linear dynamics of magnetization in ferromagnetic layers with the interface Dzyaloshinskii-Moriya interaction (IDMI). We show that IDMI leads to pinning of dynamic magnetization at the interface. An unusual peculiarity of the IDMI-based pinning is that its scales as the spin-wave wave number. We incorporate these boundary conditions into an existing numerical model for the dynamics of the Damon-Eshbach spin wave in ferromagnetic films. IDMI affects the dispersion and the frequency non-reciprocity of the travelling Damon-Eshbach spin wave. For a broad range of film thicknesses $L$ and wave numbers the results of the numerical simulations of the spin wave dispersion are in a good agreement with a simple analytical expression which shows that the contribution of IDMI to the dispersion scales as $1/L$, similarly to the effect of other types of interfacial anisotropy. Suggestions to experimentalists how to detect the presence of IDMI in a spin wave experiment are given.

## I. Introduction

The interfacial Dzyaloshinskii-Moriya interaction (IDMI) has been a subject of significant interest recently [1-7]. In Ref.[7] an attempt was made to construct a theory of spin waves in ferromagnetic films with IDMI. It has been found that this interaction may lead to significant non-reciprocity of the spin waves in these materials. Strictly speaking, the result in Ref.[7] is valid only for 1-atomic-layer (1ML) thick ferromagnetic layers, since the effective field of IDMI was treated as a bulk one, i.e. as acting with the same strength all across the film thickness, and also because the magnitude of this co-ordinate-independent effective field was assumed to be the same as the interface field ($D*k\hat{\mathbf{m}}$ in notations of Ref.[7]).

In the present work we are interested in the effect of IDMI on the thicker ferromagnetic layers ($L$>1ML, where $L$ is the film thickness). This case of "non-ultra-thin" films is more practical: these films are prospective candidates for future applications in magnonics [9], spin wave logic [10-12], and even in gas sensing [13]. Furthermore, although being not new [14,15], the problem of nonreciprocity of the Damon-Eshbach (DE) wave [16] for these technologically important films has recently attracted a lot of attention [17-23] because of its importance for a number of applications, such as microwave signal processing, measurement of spin polarisation of conduction electrons in ferromagnetic metals [17] and spin wave logic [6].

In addition, in Ref.[7] it has been supposed that the reported experimental results on the DE wave non-reciprocity might need to be re-examined keeping in mind a possible influence of IDMI on these data. Here we would like to note that all the existing results on this spin wave property have been obtained on experimental samples which are significantly thicker than 1ML. Furthermore, the ferromagnetic resonance (FMR) and spin waves in ferromagnetic films are so sensitive to surface and interface conditions that, for instance, with the FMR spectroscopy one can easily measure the strength of an interface exchange bias field for a $Ni_{80}Fe_{20}$ (Permalloy) film with a thickness as large as 60nm interfaced with a 3.5nm-thick IrMn layer [24]. Therefore, one may indeed expect an influence of the interface effect of IDMI on the spin wave dynamics in ferromagnetic layers with thicknesses much larger than 1ML.

In Section II, based on an idea by Rado and Weertman [25], we derive boundary conditions for dynamic magnetisation at the interface of a ferromagnetic layer with a non-magnetic metal which gives rise to IDMI. Previously, Soohoo [26] considered the effect of normal uniaxial surface anisotropy (NUSA) on spin waves and showed that it results in surface pinning of dynamic magnetisation. The case of the in-plane uni-directional interface anisotropy was revisited recently and its connection to the exchange bias effect was studied [24]. Treating the impact of this type of the interface anisotropy as an interface magnetization pinning effect allowed extraction of the strength of the interface pinning of dynamic magnetization from experimental FMR data on exchange-biased materials. Importantly, the values of the normal uniaxial interface anisotropy, which have a profound impact on the nonreciprocity of spin wave dispersion in the non-ultra-thin films [27], are in the range of several tenths of $mJ/m^2$, that is comparable to the value of the Dzyaloshinskii constant $D$ for which an impact of IDMI on the characteristics of domain wall motion in ultra-thin films is seen [28].

The derived boundary conditions demonstrate that IDMI induces interface magnetization pinning too. The form of the IDMI-induced pinning is different from all previously considered cases of surface/interface anisotropies: this interface field pins the *circular* components of magnetization and the pinning constants are of opposite signs for the clockwise- and counter-clockwise-rotating magnetization components. Another important peculiarity of IDMI is that the magnetization pinning and the frequency shift scale as the spin-wave wave number $k$. By default, the FMR spectroscopy is the most appropriate tool to probe the interface pinning/anisotropy. Unfortunately, it will fail to detect the presence of IDMI (in the absence of a spiral magnetisation ground state) since for the FMR experiment conditions ($k=0$) the strength of the surface magnetization pinning is precisely zero. One needs to rely on the travelling spin wave spectroscopy ($k \neq 0$) to probe the presence of IDMI.

In Section III we use the obtained boundary conditions to make numerical calculations of the Damon-Eshbach spin wave dispersion and nonreciprocity in ferromagnetic films in the presence of IDMI. The ground state of magnetization is assumed to be spatially uniform. We rely on the previously developed numerical model [22,27] which allows one to easily include any type of surface/interface boundary conditions for dynamic magnetization in the numerical code. In Section III we demonstrate that the interface magnetization pinning due to IDMI deforms dynamic magnetization profiles across the film thickness. An extra contribution to the exchange energy of spin waves which follows from this effect shifts the frequency of the spin waves in these materials. In a broad range of film thicknesses and wave numbers the results of the numerical calculations of the spin wave dispersion are in agreement with a simple analytical formula. For this range both demonstrate that the effect of IDMI on the spin wave frequency scales as $1/L$. The latter scaling law is typical for ferromagnetic films with surface/interface magnetization pinning which originates from the presence of surface/interface anisotropy.

In this section we also discuss possibilities of experimental detection of the impact of IDMI on spin waves in ferromagnetic films. Section V contains conclusions.

## II. Exchange boundary conditions for the dynamic magnetisation

To describe the magnetization dynamics we use the classical model of the linearized Landau-Lifshitz equation

$$\partial \mathbf{m}/\partial t = -\gamma \mu_0 (\mathbf{m} \times \mathbf{H} + \mathbf{h}_{eff} \times \mathbf{M}), \quad (1)$$

where $\gamma$ is the gyromagnetic coefficient and $\mu_0$ is the permeability of the vacuum. The dynamic magnetization vector $\mathbf{m}=(m_x, m_y)$ has only two non-vanishing components. The component $m_x$ lies in the layer plane and $m_y$ is perpendicular to this plane. Both are perpendicular to the static (equilibrium) magnetization vector $\mathbf{M}=M_s\mathbf{e}_z$ (which also lies in the sample plane), $\mathbf{e}_z$ is the unit vector in the $z$-direction, $\mathbf{H}=H\mathbf{e}_z$ is the applied field and $\mathbf{h}_{eff} = (\mathbf{h}_{effx}, \mathbf{h}_{effy})$ is the dynamic effective magnetic field. We assume that the ferromagnetic layer is magnetized to saturation. Hence the ground state of magnetization is spatially uniform and $\mathbf{M}$ is co-aligned to $\mathbf{H}$ everywhere inside the ferromagnetic layer.

As demonstrated in Ref. [25], if one starts with Eq.(1) and integrates over an infinitesimal volume region across the interface, the following is obtained:

$$\left(2A/M^2\right) \mathbf{M} \times \partial \mathbf{M}/\partial n + \mathbf{T}_{surf} = 0. \quad (2)$$

Here $\mathbf{M}$ represents the total magnetisation, $A$ is the exchange constant, $n$ is the direction normal to the interface ($n>0$ coincides with the direction of the $y$-axis of our frame of reference) and $\mathbf{T}_{surf}$ is the interface torque. The torque acting on the magnetisation vector is the vector product of the magnetisation vector and the interface effective magnetic field:

$$\mathbf{T}_{surf} = \mu_0 \int_{L-b}^{L} \mathbf{M} \times \mathbf{H}_{surf} \, dy, \quad (3)$$

where $\mu_0$ is the permeability of vacuum, , $y=L$ is the co-ordinate of the interface and $b$ is the thickness of the interface atomic layer. (Recall, $L$ is the thickness of the ferromagnetic layer.)

As shown in [7], the interface effective magnetic field originating from IDMI is given by

$$\mathbf{H}_{surf} = -\frac{2D}{\mu_0 M} \mathbf{e}_z \times \partial \mathbf{m}/\partial x, \quad (4)$$

where $D$ may be either positive or negative, depending on the material. We also assume that a plane spin wave of the Damon-Eshbach type propagates along the $x$ direction in the film, i.e. perpendicular to the applied field. Its wave number is $k$. This implies that $\mathbf{m}$ and $\mathbf{h}_{eff}$ scale as $\exp(-ikx)$ which results in the following expression for $\mathbf{T}_{surf}$ in the linear approximation:

$$\mathbf{T}_{surf} = -2iDbk[-\mathbf{e}_x m_x + \mathbf{e}_y m_y]. \quad (5)$$

On substituting of Eq.(5) into (2) we obtain the interface boundary conditions for the dynamic magnetisation:

$$\partial m_y / \partial y + \frac{iDkb}{A}\frac{n}{|n|}m_x = 0$$
$$\partial m_x / \partial y - \frac{iDkb}{A}\frac{n}{|n|}m_y = 0$$
, (6)

where $n$ is the *inward* normal to the interface. (This normal is directed *into* the ferromagnetic layer. For instance, $n/|n|=1$ for the layer surface (interface) $y=0$, and $n/|n|=-1$ for the layer surface (interface) $y=L$.)

Let us analyse Eq. (6). Firstly, one sees that, contrary to the boundary conditions resulting from the surface (interface) uniaxial anisotropy [26], these conditions "mix up" the $m_x$ and $m_y$ components at the interface. Indeed, the conditions in Ref. [26] are written down for each component of dynamic magnetisation separately. Conversely, each of Eqs. (6) involves both components of the magnetization vector. However, on introduction of the circular variables $m_x=(m^{(1)}+m^{(2)})/2$ and $m_y=(m^{(1)}-m^{(2)})/(2i)$ (where $i$ is the imaginary unit) the boundary conditions for vector components of the dynamic magnetisation separate:

$$\partial m^{(1)} / \partial y - d_D \frac{n}{|n|}m^{(1)} = 0$$
$$\partial m^{(2)} / \partial y + d_D \frac{n}{|n|}m^{(2)} = 0$$
, (7)

where $d_D = Dkb / A$.

This form of boundary conditions is similar to one for the dynamic magnetisation components in the Cartesian frame of reference for the case of NUSA (Eqs. 28 and 29 in Ref. [26]). The case of NUSA is well established. Therefore we may use similarity between the two cases to predict the effect of IDMI on the spin waves and FMR.

In Ref. [26] the parameter analogous to $d_D$ determines the strength of magnetization pinning at a film surface. For this reason in the following we will term $d_D$ a pinning parameter. Basically, considering surface/interface pinning of any origin, for the zero value of a pinning parameter the dynamic magnetization at the respective surface (interface) is free to precess with the same amplitude as in the bulk of the film. (This situation is often referred to as "unpinned surface spins"). The surface (interface) spins are completely pinned for the infinite value of the pinning parameter. In this situation the respective component of dynamic magnetization is zero at the interface. In a general case the pinning parameters for the two components of **m** may be quite different. In particular, for the Damon-Eshbach wave in the presence of NUSA, one component of magnetization may be completely pinned and the other one is always completely unpinned (this configuration corresponds to $\phi_{eq}=0$ in Eq.28 in [26]).

This analogy suggests that IDMI results in pinning of dynamic magnetisation at the interface. The clockwise and anti-clockwise rotating components of the dynamic magnetisation are pinned differently: the pinning constant for $m^{(1)}$ is $d_D$ and is positive for $k>0$, but the pinning constant for $m^{(2)}$ is $-d_D$ (negative for the same $k$). One also notices that the pinning scales linearly with $k$. For $k=0$ the pinning is absent completely. Hence, unfortunately, one cannot detect the presence of IDMI with the simple tool of FMR spectroscopy which is an experimental method which selectively accesses the $k=0$ point of the spin wave dispersion law. The pinning constant $d_D$ is also an *odd* function of $k$. This confirms the finding in Ref.[7] that the IDMI should lead to frequency non-reciprocity of spin waves (which is a difference in wave frequencies for $+k$ and $-k$). Interestingly, the signs of the pinning constants for the $m^{(1)}$ and $m^{(2)}$ components swap on changing the sign of $k$.

### III. Numerical simulations of spin wave spectra

We incorporate Eq.(6) into the existing numerical code [27] which solves the linearized Landau-Lifshitz Equation (1). We model a ferromagnetic layer of thickness $L$ interfaced with a non-magnetic layer. The non-magnetic layer is not included in the calculation. Its presence is taken into account by applying the IDMI exchange boundary conditions (Eqs.(6)) at the interface $y=0$. The applied field **H** and the wave vector $k$ both lie in the film plane and are perpendicular to each other (see the previous section) which forms the conditions of propagation of a Damon-Eshbach (DE)-type spin wave.

The dynamic effective field $\mathbf{h}_{eff}$ has two components: the exchange field $\mathbf{h}_{ex} = \alpha(\partial^2/\partial x^2 + \partial^2/\partial y^2)\mathbf{m}$ and the dynamic demagnetizing (dipole) field $\mathbf{h}_d$. We seek the solution of (1) in the form of a plane spin wave $\mathbf{m},\mathbf{h}_{eff} = \mathbf{m},\mathbf{h}_{eff} \exp(i\omega t - ikx)$ (see above). Therefore, the expression for the exchange field takes the form

$$\mathbf{h}_{ex} = \alpha(-k^2 + \partial^2/\partial y^2)\mathbf{m}, \quad (8)$$

where the exchange constant $\alpha = 2A/(\mu_0 M^2)$.

Similarly, the amplitude of the wave of $\mathbf{h}_d$ is given by the magnetostatic Green's function in the Fourier space $\mathbf{G}_k(s)$ [29]

$$\mathbf{h}_d(x) = \int_0^L \mathbf{G}_k(y-y')\mathbf{m}(y')dy' \equiv \mathbf{G}_k \otimes \mathbf{m}. \quad (9)$$

In our frame of reference the components of this function take the form

$$\mathbf{G}_k(s) = \begin{pmatrix} G_{kxx} & G_{kxy} \\ G_{kxy} & G_{kyy} \end{pmatrix} = \begin{pmatrix} -\delta(s) + G_p(k,s) & iG_q(k,s) \\ iG_q(k,s) & -G_p(k,s) \end{pmatrix}, \quad (10)$$

where $G_p = \frac{|k|}{2}\exp(-|k||s|)$, $G_q = \mathrm{sign}(s)\frac{k}{2}\exp(-|k||s|)$, $\delta(s)$ is Dirac delta function, and sign($s$)=1 for $s>0$ and $-1$ for $s<1$. Note that the only place where the sign of $k$ matters is the pre-factor of the expression for $G_q$. Thus, the whole information about the non-reciprocity of SW for $D=0$ is contained in the sign of this pre-factor. On substitution of (8)-(10) into (1) and introduction of the circular variables (see Section II) the linearized Landau-Lifshitz equation takes a very simple form

$$\omega \mathbf{m} = \begin{pmatrix} -[\omega_H + \omega_M(\alpha \partial^2/\partial y^2 - \alpha k^2 + 1/2)]\delta & \omega_M(G_q + G_p - \delta/2) \\ \omega_M(G_q - G_p + \delta/2) & [\omega_H + \omega_M(\alpha \partial^2/\partial y^2 - \alpha k^2 + 1/2)]\delta \end{pmatrix} \otimes \mathbf{m}, \quad (11)$$

where $\delta = \delta(s)$ (the Dirac delta function, as above), $\omega_H = \gamma \mu_0 H$, $\omega_M = \gamma \mu_0 M$ and the column vector $\mathbf{m}$ has now components $(m^{(1)}, m^{(2)})$.

One sees that the eigen-frequency of spin waves represents an eigenvalue of the integro-differential operator given by the brackets on the right-hand side of (11). Accordingly, the eigen-functions of the operator represent the modal profiles for the respective spin wave modes.

The presence of the differential parts requires application of boundary conditions at the film surfaces and interfaces. The boundary conditions are called "exchange boundary conditions" for this reason.

In the following, we solve the boundary-value problem for the integro-differential equation numerically. An alternative way of treatment of the dipole exchange spin wave dispersion problem is by introducing a scalar magnetostatic potential to describe the dipole-dipole interactions. In that case, the linearized Landau-Lifshitz equation transforms into an ordinary differential equation of 6$^{th}$ order [30]. The boundary-value problem for this equation allows analytical solution. This solution takes the form of a linear combination of six standing spin waves across the film thickness. The six wave numbers are solutions of the characteristic equation for this differential equation. The characteristic equation represents a polynomial of 6$^{th}$ order and needs to be solved numerically. Thus, ultimately, this alternative method is semi-analytical only. Furthermore, the analysis of the roots of this equation requires a significant effort [31].

Therefore, we proceed in a more established way of the direct numerical solution of the integro-differential equation [27,22]. To solve the eigenvalue problem numerically the integro-differential operator is discretized. The respective one-dimensional equidistant mesh consists of $N$ points ($j$=1,2,…$N$) located between $y$=0 and $y$=$L$. This operation transforms the equation into a matrix $C$ of a size $2N \times 2N$. The matrix's eigenvalues represent the spin wave eigen-frequencies. The eigenvectors of $C$ are spin wave mode profiles – the values $m_x(y_j)$ and $m_y(y_j)$ at the points of the mesh $y_j$. Most of the elements of $C$ do not depend on the assumed exchange boundary conditions at the layer surfaces, so they are the same for any type of surface/interface anisotropy.

The boundary conditions in the form (6) are incorporated into the discrete version of the exchange operator at the interface. To this end we use the same approach as described in Appendix 1 in Ref.[32]. The inclusion of the boundary conditions modifies the elements of $C$ for the mesh points at the vicinity of the interface. We assume that the dynamic magnetization is completely unpinned at the other surface of the ferromagnetic layer (i.e $\partial m^{(1)}/\partial y = \partial m^{(2)}/\partial y = 0$ at $y$=$L$). The IDMI boundary conditions are applied to the layer surface $y$=0. The incorporation of the boundary conditions into the block matrix $C$ results in addition of a term to the diagonal elements of its (1,1) block. This extra term reads: $2i\gamma\alpha\mu_0 M d_D/\Delta^2$ (or $2i\gamma\alpha\mu_0 M d_D/a^2$ if the mesh step $\Delta = L/N$ is equal to the lattice constant $a$). One sees that this term is an odd function of $M$, $k$ and $D$.

The eigenvalue-eigenvector problem for the matrix $C$ is solved numerically by using the QR-algorithm function built into the commercial MathCAD software. The calculation of the whole set of $2N$ eigenvalues is repeated for a number of wave numbers $k$ to produce the dispersion curve for the Damon-Eshbach mode. We are interested only in the thicknesses of ferromagnetic layers for which the effect of the exchange boundary conditions is expected to be noticeable (0-30nm). Furthermore, we are only looking at the wave number dependence of the lowest positive eigen-value of $C$. In this thickness range the frequency of the Damon-Eshbach branch of the dipole-exchange spectrum of a ferromagnetic layer is the lowest from the whole multimode spectrum (see e.g. [22]) and thus is given by the lowest eigen-value of $C$.

### IV. Discussion
#### A. Spin wave dispersion

In our computations we keep the mesh step $\Delta$ equal to the lattice constant $a$ for Permalloy: 0.3548nm. This step size choice reflects the discreteness of the real atomic lattice. Accounting for the discretness is important for simulations for thinner films. The computations are carried out for $D$=4.2 mJ/m$^2$. This value is realistic [8,28]. Given the fcc crystal structure for Permalloy (essentially nickel) the thickness of the interface atomic layer $b = a/\sqrt{2} = 0.248$ nm. This gives $Db/a$=3mJ/m$^2$.

The results of the computation for the magnetic parameters of Permalloy, $H$=300 Oe, and $L$=10$a$ (i.e. $L$=3.55 nm) are shown in Fig. 1. The applied magnetic field is *co*-aligned to the Dzyaloshinskii field ($H$>0 and $D$>0). One sees that the presence of IDMI shifts the dispersion curve upward or downward in frequency depending on the sign of $k$. The shift $\Delta f_{D,0}$=$f(D,k)$–$f(D$=0$,k)$ due to IDMI grows with $k$. This is consistent with an increase in the magnitude of the pinning constant $d_D$ with $k$.

The difference in the frequencies for $k$>0 and $k$<0 implies that the wave is characterised by frequency non-reciprocity. This is in agreement with the numerical simulations for the ultra-thin films in Ref. [7] and our analysis of the boundary conditions from the previous section. The largest spin-wave wave number which can be detected in a Brillouin light scattering (BLS) experiment [33] typically operating with a green light source is about 25 $\mu$m$^{-1}$. This is the largest $k$ value in Fig. 1. One sees that $\Delta f_{nr}$=$f(25\ \mu$m$^{-1})$–$f(-25\ \mu$m$^{-1})$=1.2 GHz. This value is significantly larger than the frequency resolution of BLS setups (100MHz). Importantly, this frequency difference is smaller than predicted by Eq.(12) in Ref.(7) by one order of magnitude.

In Fig. 2 we demonstrate $\Delta f_{D,0}$ and $\Delta f_{nr}$ as a function of $L$. The upper panel is for $|k|$=25 $\mu$m$^{-1}$ and the lower one is for $|k|$=7.8 $\mu$m$^{-1}$. The latter $k$-value is typical for the largest wave number accessible in a travelling-spin-wave spectroscopy experiment [17]. One sees that $\Delta f_{D,0}$ (upper panel) quickly decreases with an increase in the thickness. This reflects the fact that IDMI is an interface effect.

Importantly, this calculation reveals that the dependences of $\Delta f_{D,0}$ and $\Delta f_{nr}$ on 1/$L$ are close to linear and they become perfectly linear for some particular range of $L$ and $k$ values. This is illustrated in the lower panel of Fig. 2. This range is characterised by small wave numbers ($kL$<<1) and layer thicknesses >10$n_0$ or so (where $n_0$=$L/a$). The slope of this linear dependence extracted from Fig. 2 is equal to 3.32 MHz which is quite close to $2\gamma\mu_0 D^*kb/a$=3.12 MHz, where $D^*$ is defined as in Ref.[7] ($D^*$=2$D/(\mu_0 M)$ ).

Interestingly, that in this $k$ and $L$ range the spin wave dispersion is in a good agreement with a formula which is obtained by averaging **m** and **h**$_{eff}$ over the film thickness:

$$\omega(k) = \sqrt{[\omega_H + \omega_M(\alpha k^2 + 1 - P)][\omega_H + \omega_M(\alpha k^2 + P)]} + \gamma\mu_0 D^* kb/L. \quad (12).$$

In this formula the average value of the dipole field is given by the element $P(k) = 1 + \frac{1}{L}\int_0^L dy \int_0^L G_p(y-y')dy = 1 - [1 - \exp(-|k|L)]/(|k|L)$ [29,22] and the mean value of the IDMI field is $iD^*k\frac{b}{L}\mathbf{m}$.

For $D^*=0$ this formula reduces to the well-established approximate dispersion law for the Damon-Eshbach wave [29]. The last term in (12) depends on $D$. It scales as the inverse thickness, which is in agreement with our rigorous numerical result. Importantly, from this formula one obtains that $\omega(+k) - \omega(-k) = 2\gamma\mu_0 D^* kb/(an_0)$. This is in a good agreement with the plot in the lower panel of Fig. 2.

One also sees that for $n_0=1$ Eq.(12) takes the form similar to Eq.(12) in Ref.[7]. This suggests that the formalism in Ref.[7] is valid only for one-unit-cell (more precisely one-monolayer) thick ferromagnetic layers. Note that contrary to Ref.[7], where the expression for the dipole field is rather phenomenological, the dipole-dipole interaction contribution to our Eq.(12) is obtained by the mathematically rigorous procedure of averaging the dynamic variables over $L$. Thus, it is more physically sound. As previously shown [34], the method of averaging the dipole field works fine for $P<0.5$. More precisely, for $kL<0.96$ the difference between the approximate dispersion law and the rigorous Damon-Eshbach formula is 8% and for $kL<0.06$ it drops below 1%. The discrepancy is related to the increase in the surface character of the Damon-Eshbach wave. The surface character is governed by $G_q$ in Eq.(10) (see e.g. [22]). Due to its anti-symmetric character the contribution of $G_q$ to Eq.(12) is averaged out. This results in the increase in the error of the approximate expression with an increase in $k$. For $k=7.8\mu m^{-1}$ and $L=100a$ ($kL=0.03$), the error is 0.5%. Therefore Eq.(12) works fine for this $k$ value. For larger $k$-values, e.g. for $2.5\mu m^{-1}$ the surface character of the wave becomes significant for larger $L$. As a result, the approach of averaging the dipole and IDMI fields fails and one has to rely on numerical simulations.

Furthermore, our numerical simulations show that $|\Delta f_{D,0}|$-values are different for $D>0$ and $D<0$ for a given sign of $H$. This difference becomes significant for large $k$-values. In particular, in our example of $|k|=25$ $\mu m^{-1}$ and $|D|b/a=3mJ/m^2$ $\Delta f_{+D,0}=+13$ MHz and $\Delta f_{-D,0}=-16$ MHz for the positive $k$. For the negative $k$, $\Delta f_{+D,0}=-86$ MHz and $\Delta f_{-D,0}=+81$ MHz. This effect is absent in Eq.(12) which suggests that it is related to the surface character of the wave and the fact that IDMI is an interface effect.

We will elaborate on this below while considering the modal profiles of the waves. Now we only note that our calculations show that the spin wave dispersion obeys the following symmetry laws:

$$f(D,k,H,0) = f(D,-k,-H,0), \quad (13a)$$
$$f(D,k,H,0) = f(-D,k,-H,L). \quad (13b)$$

(The last index - 0 or $L$ - indicates at which film surface the IDMI boundary conditions are applied – at $y=0$ or $y=L$.) From these relations one sees that the presence of IDMI reduces the system symmetry such that the cases when both $D\mathbf{e}_z$ and $\mathbf{H}$ are aligned along $+z$ or along $-z$ are not equivalent.

Note that in reality it is not necessary to change the sign of *D* to satisfy the law (13b). It is naturally satisfied by flipping the experimental sample upside down by rotating it by 180 degree around the direction of wave propagation (*y*). Flipping the sample does not change the sign of *D*, since the latter is a physical property of a particular sample. However, this operation changes the directions of $D\mathbf{e}_z$ and **H** to the opposite ones while keeping the direction of the wave vector the same. An important consequence of the symmetry property (13b) is that its demonstration by our software implies that our numerical code is consistent.

### B. Modal profiles

Inspecting the distributions of dynamic magnetization across the thickness of the ferromagnetic layer ("modal profiles") clarifies the origin of the found frequency nonreciprocity. In accordance to Eq.(7), in Fig. 3 we plot the distributions of $m^{(1)}$ and $m^{(2)}$. For this figure we use an unrealistically large value of $|D|$=42 mJ/m$^2$ in order to accentuate the changes to the profiles IDMI introduces.

For *D*=0 (Fig. 3(a)) the larger component – $m^{(1)}$ – is characterized by an almost uniform distribution of amplitude. The smaller component – $m^{(2)}$ – is asymmetric across the thickness. This reflects the surface character of DE wave: in Fig. (3(a)) the wave propagating in the positive direction of the *x*-axis (*k*>0) is localized at the film surface *y*=*L* and the wave propagating in the opposite direction (*k*<0) is localized at the film surface *y*=0. This type of wave localization is anomalous; the wave is localized at the surface opposite to the one of localization of the exchange-free Damon-Eshbach waves [16]. As shown in [22], the anomalous localization is typical for thin metallic films.

For *D*=+42 mJ/m$^2$ one sees an increase in the interface pinning for the larger magnetization component – $m^{(1)}$ – for *k*>0: at *L*=0 $m^{(1)}$ is noticeably smaller than for *D*=0. Conversely, for *k*<0 the component $m^{(1)}$ at *y*=0 is larger than for *D*=0 which implies that the interface pinning for *k*<0 is negative. This is consistent with Eq.(7) from which one sees that the values of the pinning constant $d_D$ swap on the change of the direction of wave propagation.

Interestingly, from the comparison of Panels (b) and (c) with Panel (a) one notices that the $m^{(2)}$ component is affected only weakly by the presence of IDMI. Since from Fig. 3(a) it follows that the surface character of the Damon-Eshbach wave is mostly concentrated in the $m^{(2)}$-component, the strong similarity of $m^{(2)}$-traces for all three panels suggests that the surface character of this component is mostly determined by the dipole field; the interface IDMI field has only a minor effect on it.

An opposite tendency is visible for a negative *D*: the amplitude of $m^{(1)}$ for *k*>0 (*k*<0) at the interface is larger (smaller) for *D*=−42 mJ/m$^2$ than for *D*=0. This suggests that $m^{(1)}$ is now characterised by negative (positive) interface pinning for *k*>0 (*k*<0). Again, this is in agreement with Eq.(7) which shows that the sign of $d_D$ swaps on the change in the sign of *D*. One also notices that the profile of $m^{(1)}$ for (*D*>0,*k*>0) is practically identical to one for (*D*<0,*k*<0). However, one notices a visible difference in the *shape* of the $m^{(2)}$-profiles in the three panels: the profile in Panel (b) is the curviest one and the one in (c) is the most linear. This explains the above-mentioned fact that *f*(*D*,*k*,*H*,0)≠*f*(−*D*, −*k*,*H*,0): this fact is a joint effect of the dipole and IDMI contributions to the wave energy. The dipole-dipole interaction breaks the symmetry of the system by forming the surface-like modal profiles. The presence of IDMI field only at one of the film surfaces breaks the symmetry further. The IDMI contribution is odd in *D* but the dipole-dipole one is even (i.e. independent from *D*). Furthermore, both are odd in $k$. As a result, the $m^{(2)}$-profiles jointly affected by the dipolar field and IDMI have noticeably different shapes for ($D>0, k>0$) and ($D<0, k<0$). The difference in the shapes results in a difference in frequencies for the two cases.

This effect is similar to the effect of other types of surface anisotropies on the spin wave dispersion. In particular, the presence of NUSA only at one of the film surfaces also results in frequency nonreciprocity [15] for the Damon Eshbach wave. Similarly, it can be explained by deformation of the modal profiles. Furthermore, the case of IDMI has similarity to the case of a symmetry break by a thickness non-uniformity of the internal static magnetic field in the sample [27], which also leads to frequency nonreciprocity.

### C. Amplitude non-reciprocity

In Ref. [7] it has been pointed out that potentially IDMI is able to affect the amplitudes of excitation of spin waves by microstrip transducers (antennas) in the travelling spin wave spectroscopy experiment. It may modify the excitation-amplitude nonreciprocity and the existing literature results must be reconsidered accordingly. The pertinent experiments are [17,18,21,23]. All of them have been conducted on ferromagnetic films with $n_0>1$.

To check this claim we evaluate the excitation-amplitude nonreciprocity, based on the ideas from [20] and [35]. The Fourier component $\mathbf{h}_{k\,exc}$ of the microwave magnetic field of the stripline antenna is given by the equation as follows: $\mathbf{h}_{k\,exc}=(\mathbf{e}_x-i\mathbf{e}_y)j_k\equiv|1,-i\,\text{sign}(k)>j_k$, where $j_k\mathbf{e}_z$ is the Fourier component of the microwave current density in the antenna (see e.g. Eq.(15) in [20]). The scalar dimensionless amplitude $A_k$ of the excited DE wave scales as $<\mathbf{m}|1,-i>j_k$, where $<\mathbf{m}|\equiv<m_x, m_y|$ is the respective left-hand eigenvector of the matrix $C$, $<\mathbf{m}|\mathbf{m}>=1$, and $|\mathbf{m}>$ is the right-hand eigenvector of $C$. ($<\ldots|\ldots>$ denotes a scalar product of a pair of vectors). As a result, the ratio $R$ of the amplitudes of the waves propagating in the opposite directions from the antenna is given by

$$R = A_{-|k|} / A_{+|k|} = <\mathbf{m}(-|k|)|1,+i> / <\mathbf{m}(+|k|)|1,-i>. \quad (14).$$

For $R=1$ the wave is fully reciprocal and for $R=0$ or infinity the wave excitation is unidirectional. In Fig. 4 we plot $R$ for $D=0$ and $Db/a=\pm3$ mJ/m$^2$. Similarly to the lower panel of Fig. 2, we use the range of spin-wave wave numbers typically accessible in the travelling spin wave spectroscopy experiment: from 0 to 7.8 μm$^{-1}$. One sees that IDMI modifies $R$: for $D<0$ ($D>0$) $R$ is larger (smaller) than for $D=0$. However, the effect is rather small. Thus, it will be hardly possible to study it experimentally.

Note that one has to keep in mind that in this graph we show the $k$-dependence of $R$. The $f$-dependence of $R$ (not shown) will also include a contribution from the frequency nonreciprocity (Fig. 2). This is illustrated by thin lines in Fig. 4 which demonstrate $\Delta f_{nr}$ for $Db/a=\pm3$ mJ/m$^2$.

### D. Implications for future experiments

We begin this sub-section by discussing the effect of IDMI on the ground state of magnetization. Our theory is valid for the spatially uniform state of static magnetization. Therefore it is important to understand whether for a particular set of system parameters the uniform ground state exists, or a non-uniform ground state characterised by a spiralling magnetization vector is formed instead. We may try to estimate the possibility of formation of the spatially non-uniform magnetization ground state by looking at the possibility of softening of spin wave dispersion. This analysis can be done based on Eq.(12). From this equation, one sees that for $Dk<0$ the spin wave frequency may become zero. The zero frequency ("mode softening") may be considered as an indication of possibility of formation of a spatially periodic ground sate. The value of k for which the frequency becomes zero may be considered as a proxy to the characteristic period $2\pi/k$ of the periodic ground state.

From Eq.(12) one sees that the contribution of the exchange energy to the spin wave frequency scales roughly as $\omega_M \alpha k^2$, i.e. as a square of the wave number. Thus, it is always positive, even in $k$, and represents quite a steep function. The IDMI contribution $-D*kb/L-$ is linear and odd in $k$. The outcome of the competition of the two contributions depends on particular parameters of the sample and the experiment. For $D*kb/L<0$ it may happen that for smaller $k$ values $f$ decreases with an increase in $k$, because of the dominance of the linear term. However, for larger $k$ values the contribution of the quadratic term will kick in and the frequency will start to grow inevitably. Whether the frequency is able to drop all the way to zero or not depends on $L$ in the first place. For a 3.55nm-thick film there is no section of the dispersion curve with a negative slope, as one sees from Fig. 1. This implies that the ground state for a film this thin is highly likely to be spatially uniform and the totality of the analysis of spin wave dispersion from Subsections IVA-IVC should be valid for this film thickness. However, our numerical calculations show that for smaller film thicknesses (e.g. $n_0$=3) negative dispersion is possible. As follows from Eq.(12), whether the frequency drops to zero or not, depends on the applied field: by increasing $H$ one can always achieve the situation when the minimum frequency is larger than zero and thus the uniform ground state is stabilized.

Two types of experiments are typically used to probe the spin wave dispersion in thin ferromagnetic metallic films: BLS [33] and stripline-antennae based travelling spin wave spectroscopy (TSWS) [36]. The maximum spin-wave wave number which can be detected with a BLS setup operating with a green light is 25 $\mu m^{-1}$. The respective frequency resolution is 100 MHz or so. Importantly, the sensitivity of the BLS setups is sufficient for characterising films with thicknesses down to a couple of atomic layers.

The frequency resolution of the stripline antenna based TSWS spectrometers is much better: the change in the frequency of a fraction of MHz can be easily detected [17,27]. However, the maximum spin-wave wave number detected so far with a stripline transducer is significantly smaller: 7.8 $\mu m^{-1}$. This is limited by the capabilities of the modern lithography to define the antenna geometry and also largely by the necessity of a good impedance match of the microscopic antenna to the external microwave circuit. Furthermore, the spin wave group velocity scales as 1/$L$. This scaling results in a very small free propagation path for spin waves - just a couple of microns - in the films with $L$<10nm. This makes the output microwave signal of the structure containing the film and the two antennas become comparable to the noise level.

From both panels of Fig. 1 one sees that there is a good chance to detect the presence of IDMI with both BLS and TSWS. There might be two ways to study the effect of IDMI on the spin wave dispersion. The first one is by fabricating a pair of samples ("reference sample method"). One is a single-layer ferromagnetic film which will serve as a reference sample "$D$=0". The second sample is a bi-layer film with the same ferromagnetic layer but interfaced with a non-magnetic layer which presumably induces IDMI in the ferromagnet. Then one can measure the differences in the spin wave dispersions (with either TSWS or BLS) and in the excitation amplitudes (TSWS), or in the BLS intensities.

However, this is not the cleanest way to set up an experiment, because the reference film may spontaneously develop NUSA which will result in magnetization pinning at the film surface and compromise the comparative study. Therefore, it would be better to avoid using a reference sample. To implement this reference-sample-free protocol one will need to measure relative changes in the sample response as a function of experiment parameters and to infer about the presence of IDMI from the form of these dependences. The experiment may be set similar to the measurements carried out in [17,27]. One takes four measurements of spin wave frequency in total ("4-measurement method"): $f(+k,+H)$, $f(-k,+H)$, $f(-k,-H)$, and $f(+k,-H)$. A difference in $f(+k,+H)$ and $f(-k,+H)$ and equivalence of $f(+k,+H)$ and $f(-k,-H)$ will confirm the presence of IDMI. Furthermore, from the difference $f(+k,+H)-f(-k,+H)$ one will be able to extract the value and the sign of $D$.

Given the frequency resolution, as follows from Fig. 2, thinner films may be probed by BLS and thicker films are more suitable for TSWS characterisation. The amplitude non-reciprocity and its equivalent for BLS – the Stokes/anti-Stokes asymmetry of BLS intensities may be a small issue, which may make the signal of the wave propagating in the unfavourable direction weaker. However, as shown in Ref.[17] it is possible to successfully use the 4-measurement method in the TSWS experiment on films with thicknesses as small as 10nm.

### V. Conclusion

In this work we derived the interface exchange boundary conditions for the linear dynamics of magnetization in non-ultra-thin (1 atomic layer+) ferromagnetic films with the interface Dzialoshinskii-Moryia interaction (IDMI). We incorporated these boundary conditions into our numerical model for the dynamics of the Damon-Eshbach spin wave in thin ferromagnetic films based on the linearized Landau-Lifshitz Equation. Our analysis of the boundary conditions and numerical simulations demonstrated that IDMI results in an interface pinning of dynamic magnetization. An unusual peculiarity of the IDMI-based pinning is that its scales as the spin-wave wave number. As a result, no impact of IDMI will be seen in the ferromagnetic resonance spectroscopy data. One will need to rely on travelling spin wave spectroscopy experiment to detect the presence of IDMI in a sample.

IDMI affects the dispersion and the frequency non-reciprocity of the travelling Damon-Eshbach spin wave. For a broad range of film thicknesses $L$ and relatively small wave numbers the results of the numerical simulations of the spin wave dispersion are in a good agreement with a simple analytical expression which shows that the contribution of IDMI to the dispersion scales as $1/L$, similarly to the effect of other types of interfacial anisotropy. This contribution is large enough in order to allow detection of IDMI in a Brillouin light scattering and travelling spin wave spectroscopy experiments for a broad range of thicknesses of ferromagnetic layers. Suggestions to experimentalists are given how to implement those studies.

It has also been shown that there is an impact of IDMI on the amplitudes of excitation of Damon-Eshbach waves by stripline antennas. However, this contribution is small with respect to the intrinsic amplitude non-reciprocity of the Damon-Eshbach wave. Therefore it will be difficult to detect the contribution of IDMI to the amplitude non-reciprocity in the travelling spin wave spectroscopy experiment.

**Acknowledgment**

Financial support by the Australian Research Council is acknowledged.

**Figure captions**

Fig. 1. (a) Damon-Eshbach spin wave dispersion $f(k)$ for $Db/a=+3$ mJ/m$^2$. Thick solid line: $k>0$; dashed line: $k<0$. Thin solid line: the same, but for $D=0$ (given here for comparison). (b) Frequency difference $\Delta f_{D,0}$. Solid line: $k>0$. (This is the difference between the thick and the thin solid lines from Panel (a)). Dashed line: $\Delta f_{D,0}(k<0)$ (the difference between the dashed and the thin solid lines from Panel (a)). The thickness of the ferromagnetic layer equals to 10 unit cells for Permalloy ($L=10a=3.55$ nm), applied field $H=+300$ Oe, saturation magnetization $4\pi M=10.5$ kOe ($\mu_0 M=1.05$ T), exchange constant $A=1.355\times10^{-6}$ erg/cm ($1.355\times10^{-11}$ J/m). Gyromagnetic coefficient is 2.8 MHz/Oe. The film is magnetized to saturation along the $z$ direction and the equilibrium magnetization vector is co-aligned with **H**. IDMI is present at one ferromagnetic layer surface only. Dynamic magnetization at the second ferromagnetic layer surface is unpinned ($\partial \mathbf{m}/\partial y=0$).

Fig. 2. Upper panel: Frequency shift due to IDMI ($\Delta f_{D,0}$) for $k=+25$ μm$^{-1}$ (solid line) and $k=-25$ μm$^{-1}$ (dashed line) as a function of the thickness $L$ of the ferromagnetic layer. Lower panel: frequency nonreciprocity $\Delta f_{nr}$ for $|k|=7.8$ μm$^{-1}$ as a function of $1/L$.
The film thickness is given in the units of the number of unit cells of the crystal lattice: $n_0=L/a$. In the upper panel the thickness range spans from 1 nm to 35.5 nm. In the lower panel it is from 7.1 nm to 35.5 nm. All other parameters are the same as for Fig. 1.

Fig. 3. Modal profiles for the Damon-Eshbach wave in the presence of IDMI. (a) IDMI is absent ($D=0$). (b) $D=+42$ mJ/m$^2$. (c) $D=-42$ mJ/m$^2$. The layer interface with IDMI is located at $y=0$. The spins at the second surface of the ferromagnetic layer $y=L=35.5$ nm are unpinned. The other parameters are the same as for Fig. 1. Solid lines: $k=+25$ μm$^{-1}$; dashed lines: $k=-25$ μm$^{-1}$. The two upper plots in each panel are for $m^{(1)}$. The two lower ones are for $m^{(2)}$.

Fig. 4. Ratio of the amplitudes of the Damon-Eshbach spin waves excited by a microwave microstrip transducer in two opposite directions from the transducer (left-hand axis). Thick solid line: $D=0$. Thick dashed line: $Db/a=+3$ mJ/m$^2$. Thick dash-dotted line: $Db/a=-3$ mJ/m$^2$. Film thickness is $L=35.5$ nm. The other parameters are the same as for Fig. 1. Thin lines are the respective frequency non-reciprocities $\Delta f_{nr}$, given here for comparison (right-hand axis). Thin dashed line: $\Delta f_{nr}$ for $Db/a=+3$ mJ/m$^2$; thin dash-dotted line: $Db/a=-3$ mJ/m$^2$. The wave number range shown here is typical for the travelling spin wave spectroscopy (TSWS) experiment.

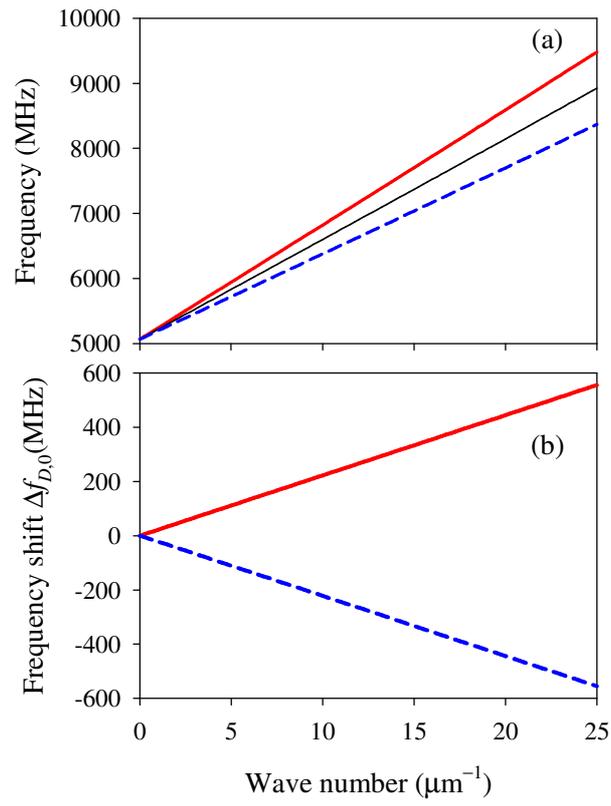

Fig. 1

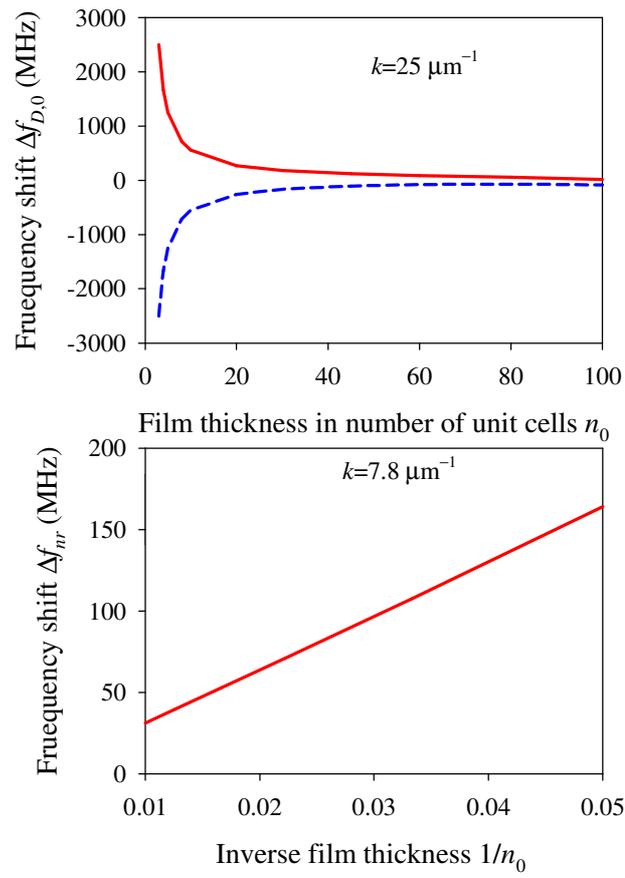

Fig. 2

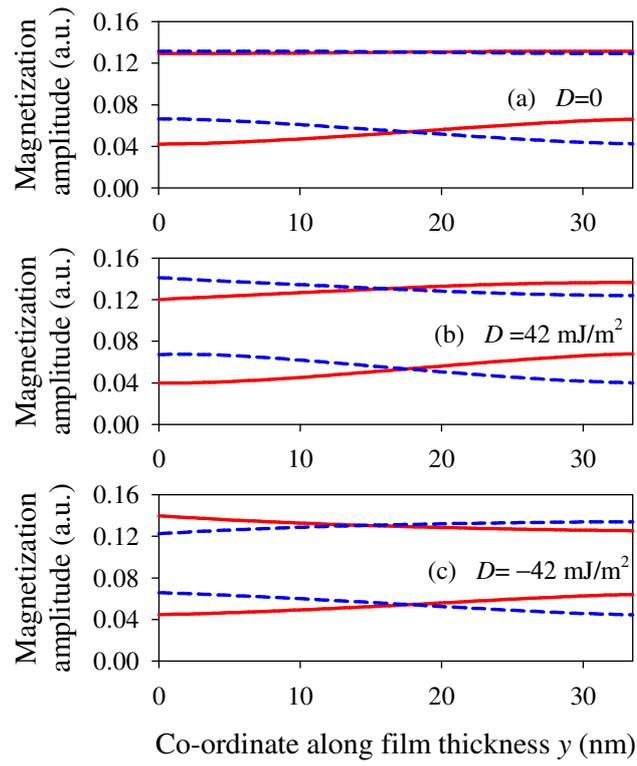

Fig. 3.

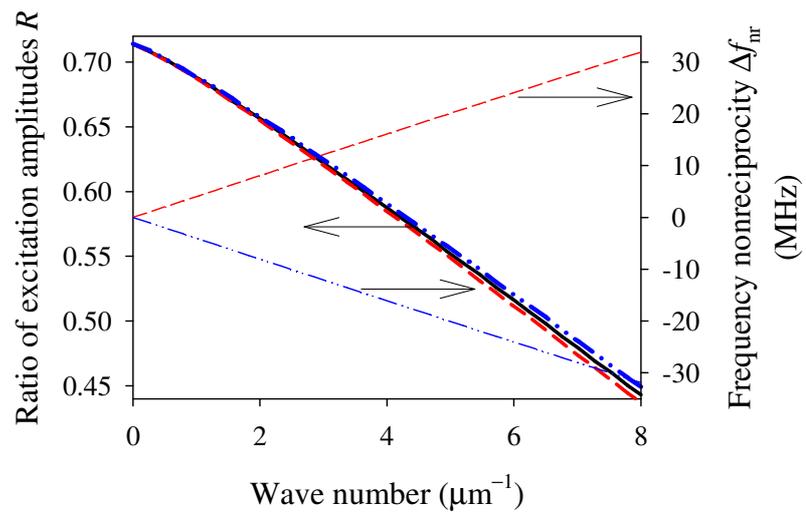

Fig. 4.